\documentclass[twocolumn,pra,superscriptaddress]{revtex4-1} 
\usepackage{graphicx}
\usepackage{amssymb,amsmath}
\usepackage{bm}
\usepackage{dcolumn}
\usepackage{subfigure}
\usepackage[dvipdfmx]{hyperref}
\hypersetup{backref,pdfpagemode=FullScreen,colorlinks=true,breaklinks,urlcolor=blue,linkcolor=blue,citecolor=blue}
\usepackage{color}
\usepackage{ulem}

\begin{document}

\title{Persistent atomic spin squeezing at the Heisenberg limit}
\author{Ling-Na Wu}
\affiliation{State Key Laboratory of Low Dimensional Quantum
Physics, Department of Physics, Tsinghua University, Beijing 100084, China}
\author{Meng Khoon Tey}
\author{L. You}
\email{lyou@mail.tsinghua.edu.cn}
\affiliation{State Key Laboratory of Low Dimensional Quantum
Physics, Department of Physics, Tsinghua University, Beijing 100084, China}
\affiliation{Collaborative Innovation Center of Quantum Matter, Beijing 100084, China}

\date{\today}

\begin{abstract}
Two well-known mechanisms, one-axis twisting (OAT) and two-axis counter twisting (TACT),
generate spin squeezed states dynamically.
The latter provides better spin squeezing (SS), but has not been demonstrated
as the form of its interaction does not occur naturally in known physical systems.
Several proposals for realizing effective TACT transformed from OAT
require stringent experimental conditions, in order to
overcome the resulting non-stationary (oscillating)
SS and continuously varying mean spin directions.
This work presents a simple scheme that solves both problems by freezing
SS at an optimal point and realizing effectively persistent SS by
inhibiting further squeezing dynamics. Explicit procedures are outlined for
persistent SS of the TACT limit. Protocols based on
our scheme favorably relax experimental demands, which significantly brighten
the prospects for realizing TACT.
\end{abstract}

\maketitle

Recent successes in atomic spin squeezing (SS) have significantly raised the prospects
for its application to high precision measurement \cite{PhysRevA.46.R6797,*PhysRevA.50.67,sorensen2001many,leibfried2004toward,gross2010nonlinear,riedel2010atom,PhysRevLett.104.073602,Polzik2010,giovannetti2011advances,Nonlinear2012,PhysRevLett.109.253605,PhysRevA.90.062132,bohnet2014reduced,hamley2012spin}
and to entanglement detection \cite{PhysRevLett.86.4431,PhysRevLett.95.120502,PhysRevA.74.052319,PhysRevA.79.042334,PhysRevLett.102.100401,PhysRevA.86.012337}.
Squeezed spin state (SSS) \cite{PhysRevA.47.5138,PhysRevA.46.R6797,*PhysRevA.50.67,ma2011quantum}
is a state of many spin (pseudo-spin) 1/2 particles with exchange symmetry,
whose uncertainty in one collective
spin component perpendicular to the mean spin direction is smaller than
the classical limit set by coherent spin state (CSS), where all spins are
identically aligned up in the same direction.
Kitagawa and Ueda \cite{PhysRevA.47.5138} proposed two well-known mechanisms
for generating SS with one-axis twisting (OAT) and two-axis counter twisting (TACT) interactions,
which are described respectively by the Hamiltonians ${H_{{\rm{OAT}}}} = \chi J_z^2$
and ${H_{{\rm{TACT}}}} = \chi \left( {J_z^2 - J_y^2} \right)$.
Here ${J_i} = \sum\nolimits_k {\sigma _i^{(k)}/2}\, \left( {i = x,y,z} \right)$ denote
the collective spin components, where $\sigma _i^{(k)}$ are the
Pauli matrices for the k-th spin 1/2 particle,
and $\chi$ denotes the identical strength of coupling between two spins.
The squeezing parameter ${\xi ^2} = \left( {\Delta {J_ \bot }} \right)_{\min }^2/\left( {\Delta {J_ \bot }} \right)_{\rm CSS}^2$
quantifies the degree of SS in terms of the minimum of the fluctuation
$\left( {\Delta {J_ \bot }} \right)^2 = \langle J_ \bot ^2\rangle - \langle {J_ \bot }\rangle ^2$
for the collective spin component perpendicular to the mean spin
$\langle \vec J\rangle  = \left( \langle {J_x}\rangle ,\langle {J_y}\rangle ,\langle {J_z}\rangle\right)$
relative to $\left( {\Delta {J_ \bot }} \right)_{\rm CSS}^2=N/4$ for a CSS with $N$ the number of particles.

The theoretical limits of the squeezing parameters for OAT and TACT scale as
$\propto {N^{-2/3}}$ and $\propto N^{-1}$, respectively \cite{PhysRevA.47.5138}.
Despite its better scaling capable of approaching
within a few times of the Heisenberg limit $(1/N)$, TACT is yet to be demonstrated as its form of
interaction does not occur naturally in most systems of interest for studying SS.
OAT, on the other hand, has been implemented in many systems \cite{orzel2001squeezed,esteve2008squeezing,gross2010nonlinear,riedel2010atom,PhysRevLett.101.073601,PhysRevLett.114.043604}.
Many theoretical studies have been carried out to realize TACT interactions \cite{PhysRevLett.107.013601,huang2014two,PhysRevLett.87.170402,PhysRevA.65.053819,PhysRevA.65.041803,PhysRevA.68.043622,PhysRevA.91.053612,PhysRevA.90.013604}.
Two promising proposals realize TACT by coherent manipulation of OAT \cite{PhysRevLett.107.013601, huang2014two}.
Specifically, in Ref. \cite{PhysRevLett.107.013601}, an alternating $\pi/2$ pulse train
periodically switches the OAT axis; and
in Ref. \cite{huang2014two}, a periodically modulated drive is applied
to continuously modify the direction of atomic spin.
Although TACT SS is realized in both cases at integer periods of modulation,
the amount of SS oscillates in time with an amplitude determined by
the degree of how well the approximation conditions adopted \cite{PhysRevLett.107.013601, huang2014two}
are satisfied.
For instance, in the formal proposal \cite{PhysRevLett.107.013601},
$1000$ or more pulses are required for a condensate with $N=1250$ atoms
in order to reach a squeezing parameter (${\xi ^2}$) within twice the limit of TACT.
Without impeccable precisions, however, errors from repeated pulses will accumulate to spoil
the intended dynamic spin control.
In the latter proposal \cite{huang2014two},
modulation frequencies as high as $10^5$ times the coupling strength $\chi$
and Rabi frequencies of the same order of magnitude are required at the same $N$.
Additionally the continuously nutating mean spin direction
from modulated drive complicates its detection.
Therefore, schemes capable of effectively suppressing SS (${\xi ^2}$) oscillation amplitude
 and tracking the varying mean spin direction
would significantly establish the proposals' feasibilities.

This paper presents an idea which solves both afore-mentioned challenges
in the two proposals \cite{PhysRevLett.107.013601, huang2014two}.
Its application freezes squeezing dynamics at the theoretically determined optimal point,
essentially achieves persistent TACT squeezing. The following discussion
starts with a description of our basic idea.
Explicit operation protocols are then provided for the two afore-mentioned
proposals \cite{PhysRevLett.107.013601,huang2014two},
which lead respectively to more than one order of magnitude reduction
in the required number of pulses and
five times reduction in the modulation (and Rabi) frequency.

We note that
a many particle state with a sharp distribution around a system eigenstate
 is less sensitive to dynamic evolution \cite{PhysRevLett.99.170405}.
For example, we consider the
OAT model ${H_{{\rm{OAT}}}} = \chi J_z^2$, whose eigenstates are
$\left\{ {\left| {j,m} \right\rangle } \right\}$,
satisfying ${J_z}\left| {j,m} \right\rangle  = m\left| {j,m} \right\rangle $, with $m=-j,-j+1,...,j$.
The eigenstate $\left| {j,m} \right\rangle$ acquires a phase $\chi t{m^2}$ after time $t$.
A state with a wide distribution of $m$ phase diffuses \cite{PhysRevLett.44.1323} quickly.
A state with a narrow distribution around a particular eigenstate ${\left| {j,k} \right\rangle }$,
however, diffuses slowly and essentially acquires a global phase factor $\chi t k^2$.
When $k\sim 0$, its temporal evolution becomes effectively frozen.

The above discussion illustrates the basics of
a robust storage scheme for SS \cite{PhysRevLett.99.170405}
in the model of OAT with a constant drive \cite{PhysRevA.63.055601}, whereby the spin distribution is in continuous rotation around the mean spin direction while undergoing SS.
The drive is turned off at precisely the moment when optimal SS lies along the $z$-axis \cite{PhysRevLett.99.170405}.
The corresponding SSS becomes insensitive
to subsequent dynamical evolution due to its squeezed distribution ($k\sim 0$) along the $z$-axis.
Instead of waiting passively for the optimal moment,
 our idea is on the active side.
We propose to rotate the maximally squeezed direction of a SSS to
along the $z$-axis instantaneously when optimal squeezing occurs, which freezes optimal SS and keeps it persistent.
In order to precisely carry out the rotation, we need to know the optimal squeezing direction.
Likewise, to reach optimal SS, an accurate knowledge of the optimal squeezing time is required.
Fortunately, these information are known for both OAT and TACT models,
e.g., the optimal squeezing time for OAT and TACT are $\chi t_{{\rm{opt}}}^{{\rm{(OAT)}}} \simeq {6^{1/6}}{N^{ - 2/3}}$ \cite{PhysRevA.47.5138} and $\chi t_{{\rm{opt}}}^{{\rm{(TACT)}}} \simeq \ln \left( {4N} \right)/\left( {2N} \right)$ \cite{PhysRevLett.107.013601}, respectively.
Although both depend on atom number $N$, this dependence is very weak
for the squeezing parameter ${\xi ^2}$ in the vicinity of the maximal squeezing,
which greatly relaxes the required precision of temporal controls.
The optimal squeezing direction is fixed in the TACT model,
independent of $N$ or other parameters, which is an advantage over the OAT model.

\begin{figure}
\centering
  \includegraphics[width=0.95\columnwidth]{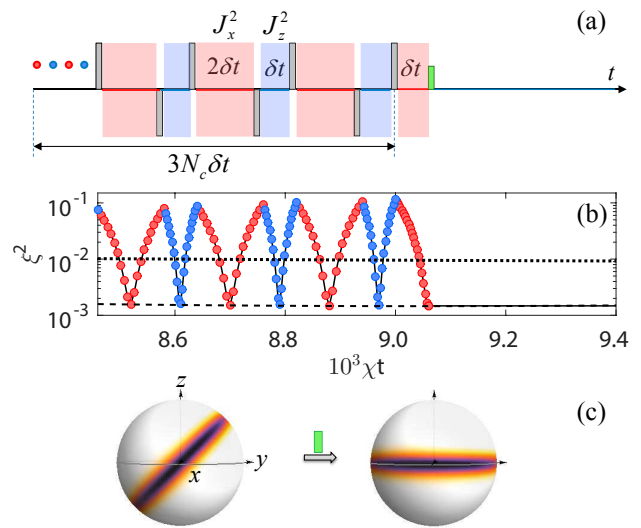}
  \caption{(Color online) (a) An illustration of the repeated pulse proposal \cite{PhysRevLett.107.013601},
   where each period lasts for $t_c=3\delta t$ and is composed of a $2\delta t$ section (red shadow)
   governed by $H_1 = \chi J_x^2$
   and a $\delta t$ section (blue shadow) governed by $H_2 = \chi J_z^2$.
   $H_1$ is transformed from $H_2$ by a pair of $\pi/2$ pulses (gray rectangles) applied along the $\pm y$-axis.
   The green rectangle represents the rotation pulse proposed
   to freeze spin evolution. (b) The evolution of squeezing parameter $\xi^2$ as a function of time.
   Red (blue) disks denote the first (second) part of duration $2\delta t$ ($\delta t$).
   Black solid line represents the result upon implementation of the present scheme.
   For all figures in this work, black dashed line and black dotted line refers to the limits of TACT and OAT models, respectively.
   (c) A Bloch sphere illustration of the rotation operation conducted to freeze SS.
   The quasi-probability distribution for a state $\left| {\psi (t)} \right\rangle $
   is $|\langle \theta ,\varphi |\psi (t)\rangle {|^2}$, where $|\theta ,\varphi \rangle$
   denotes CSS pointing along $(\theta,\varphi)$ direction.
   The initial state used for our scheme is ${\left| {j,j} \right\rangle }$, while the
   limits of OAT and TACT in all figures in this work are obtained with their appropriate CSS as initial states.
   Likewise, in this and all other figures shown later,
   $N=1250$ is used, except for illustrations with Bloch spheres where $N=100$ is used
    as in (c) for enhanced details of the quasi-probability distribution, and the
   number of periods $N_c=50$. }
  \label{fig1}
\end{figure}

The following discussions detail the operation protocols
for our scheme when applied to the two afore-mentioned proposals \cite{PhysRevLett.107.013601, huang2014two}.
First, consider the repeated pulse proposal \cite{PhysRevLett.107.013601},
 where a $\pi/2$ pulse train
periodically switches the OAT axis [Fig. \ref{fig1}(a)]. Each period lasts for $t_c = 3\delta t$ and
is composed of two parts: an evolution of $2\delta t$ (red shadow) governed
by OAT Hamiltonian $H_1 = \chi J_x^2$, followed by OAT  $H_2 =H_{\rm{OAT}}$
for the remaining $\delta t$ (blue shadow). $H_1$ is transformed
from $H_{\rm{OAT}}$ by a pair of $\pi/2$ pulses (gray rectangles) applied along the $\pm y$-axis.
Although they do not commute, i.e. $\left[ {{H_1},{H_2}} \right] \ne 0$,
provided $2\chi\delta t N\ll 1$, we can neglect higher order terms in $\delta t$
and end up with an effective TACT Hamiltonian $H_{\rm{eff}} = (2H_1+H_2)/3=\chi(J_x^2-J_y^2)/3$
apart from a constant $J^2=j(j+1)$. As shown in Fig. \ref{fig1}(b),
SS from the actual pulse sequence reaches the TACT limit (black dashed line) at specific times,
accompanied with oscillations from the higher order terms in $\delta t$.
The accuracy of this approximation can be improved by reducing $\delta t$,
which shortens $t_c$ and inhibits oscillation amplitudes. However, this is not always a winning strategy
as the optimal squeezing time $t_{\rm{opt}}$ is fixed by $N$.
A smaller $t_c$ indicates a larger $N_c$, which denotes the nearest integer number
of pulse pairs needed to reach optimal SS.
It was found earlier that more than $N_c=1000$ pairs
of pulses are needed to reach the effective TACT at $N=1250$ \cite{PhysRevLett.107.013601}.
Controlling such a large number of identical pulses to the required
accuracies is a serious experimental challenge.

\begin{figure}
  \centering
  \includegraphics[width=0.95\columnwidth]{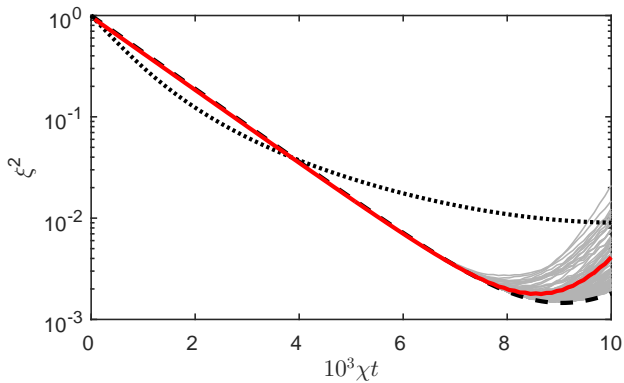}
  \caption{(Color online)
  The evolution of squeezing parameter $\xi^2$ for the repeated pulse proposal in the presence
  of added linear noise to the pulse area according to a model described in the main text.
  The grey lines show $\xi^2$ from 100 independent realizations, and the thick red line
  denotes their average.
  For easy viewing of the oscillating SS,
  only data points at $n t_c+\delta t$ and $n t_c+2.5\delta t$ for $n=0,1,2,\cdots$,
  are shown.
  }
  \label{fig2}
\end{figure}

The good news lies at the fact that
even for $N_c=50$,
despite its oscillations,
${\xi ^2}$ already touches the TACT limit at approximately $(N_c\pm n) t_c+\delta t$ and $(N_c\pm n) t_c+2.5\delta t$
for integers $n=0,1,2,\cdots$ and $N_c t_c \simeq t_{{\rm{opt}}}=3t_{{\rm{opt}}}^{{\rm{(TACT)}}}$ [Fig. \ref{fig1}(b)].
Our protocol calls for the rotation of the optimal squeezed direction to along the $z$-axis
with a short pulse [green rectangles in Fig. \ref{fig1}(a) and \ref{fig1}(c)],
at the appropriate moments when optimal SS is reached.
This freezes the optimal SS [black solid line in Fig. \ref{fig1}(b)],
which nearly overlaps with TACT limit.
The squeezing axis at maximum squeezing lies
at an angle $\pi/4$ to the $z$-axis in the $y$-$z$ plane \cite{PhysRevLett.107.013601},
which can be rotated to the $z$-axis by a $\pi/4$ pulse
along the $-x$-axis, as shown in Fig. \ref{fig1}(c).
Given $\chi\sim (2\pi)\,0.063\,$Hz as from recent experiment \cite{gross2010nonlinear} and $\chi t_{{\rm{opt}}} \sim 0.01$ for $N=1250$ [Fig. \ref{fig1}(b)],
we have $N_ct_c\sim 25\,$ms. At $N_c=50$, a single
pulse is limited to a duration $\le 10\,\mu$s as $t_c\sim 500\,\mu$s
or $\delta t \sim 170\,\mu$s, which is now feasible
in contrast to the $1000$ pulse pairs originally required.
This drastic reduction of $N_c$ also makes the proposal
more robust to a fluctuating pulse area.
Figure \ref{fig2} displays squeezing parameters $\xi^2$ from 100 independent simulations
for a fluctuating pulse area proportionally scaled to $1+r\eta$,
with $\eta=0.1\%$ and $r\in [-0.5,0.5]$ a uniformly distributed random number.
Such conditions are achievable experimentally,
and our simulations show that the quality of SS remains very good.
In particular, we note that in the broad temporal domain before the
optimal SS is reached, $\xi^2$ is found to track the limit of TACT,
essentially unaffected by the fluctuating noise and at a level significantly
below the limit of OAT.

We now apply our protocol to the modulated drive proposal \cite{huang2014two},
where a continuously modulated drive
$\Omega(t)=\Omega_0\cos \left( {\omega t + \varphi} \right)$
augments the OAT model
to give
\begin{eqnarray}\label{H_t}
H(t)=H_{\rm{OAT}}+\Omega(t) J_y.
\end{eqnarray}
In the high-frequency limit $\omega  \gg N\chi $,
this Hamiltonian at $\varphi=0$ is well approximated by
\begin{eqnarray}\label{Happ}
H_{\rm{appx}} = \chi \left[ {\alpha_0J_z^2 + \left( {1 - \alpha_0} \right)J_x^2} \right],
\end{eqnarray}
with ${\alpha_0} = \frac{1}{2}\left[ {1 + {J_0}\left( {2\Omega_0 /\omega } \right)} \right]$,
where ${J_\nu}\left(.\right)$ denotes the $\nu$-th order Bessel function,
and is bounded within $(-0.5,1]$. At $\Omega_0/\omega=0.9057$,
$\alpha_0=2/3$, Hamiltonian (\ref{Happ}) reduces to
\begin{eqnarray}\label{Heff}
H_{\rm{eff}} = \frac{\chi}{3}\left( {2J_z^2 + J_x^2} \right) = \frac{\chi}{3}\left( {{J^2} + J_z^2 - J_y^2} \right),
\end{eqnarray}
which is formally equivalent to the TACT model,
except for a constant ${J^2}{\rm{ = }}j\left( {j{\rm{ + }}1} \right)$ term.
Heisenberg limited SS can be achieved dynamically starting
from a coherent state $\left| {j,j} \right\rangle _x=\exp \left( { - i\pi{J_y}/2} \right)\left| {j,j} \right\rangle $
with all spins pointing to the $x$-direction.

In fact, this effective TACT remains applicable even for an arbitrary phase $\varphi\neq 0$
except for an extra unitary transformation (see appendix)
\begin{eqnarray}\label{hkk}
{H_{{\rm{eff}}}^\prime} = {R_y}\left( \varphi  \right)H_{\rm{eff}} {R_y}\left( { - \varphi } \right),
\end{eqnarray}
where ${R_y}\left( \varphi  \right) = {\rm{exp}}\left( { - i\frac{\Omega_0 }{\omega }\sin \varphi {J_y}} \right)$
denotes a rotation around the $y$-axis by an angle $({\Omega_0 }/{\omega })\sin \varphi$. SS at
Heisenberg limit is again realized if a rotated CSS ${R_y}\left( \varphi  \right)\left| {j,j} \right\rangle _x$ is used as the initial state.

\begin{figure}
  \includegraphics[width=0.99\columnwidth]{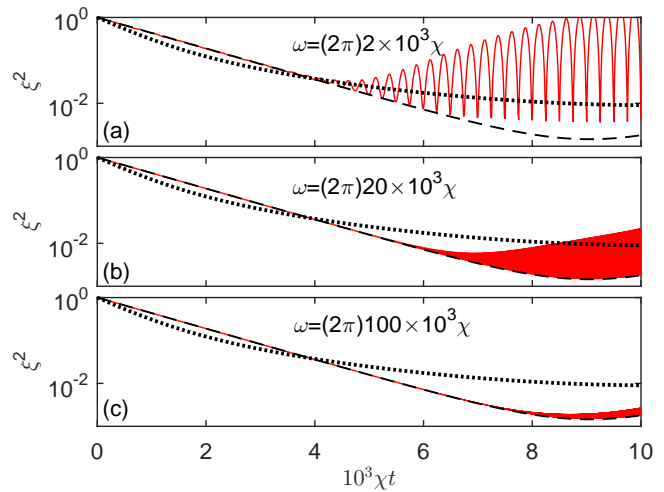}\\
  \caption{(Color online) Squeezing parameter $\xi^2$ from the actual dynamics of the Hamiltonian (\ref{H_t}) with $\varphi=-\pi/2$ (red solid lines) at $\omega=(2\pi)2\times 10^3\chi$ (a), $(2\pi)2\times10^4\chi$ (b), $(2\pi)10^5\chi$ (c) for a fixed ratio of $\Omega_0/\omega=0.9057$.
  The initial state used for the modulated drive proposal is $\exp(i\Omega_0 J_y/\omega){\left| {j,j} \right\rangle _x}$.}
  \label{fig3}
\end{figure}

Figure \ref{fig3} displays $\xi^2$ from the dynamics of the modulated drive proposal (red solid lines). Fixing $\Omega_0=0.9057\omega$, $\xi^2$ approaches and eventually
overlaps completely with results from the effective dynamics of $H_{\rm{eff}}$ (\ref{Heff}) (black dashed lines) when the modulation frequency $\omega$ increases.
To suppress oscillation amplitude to within $50\%$ of the TACT limit,
the high frequency approximation requires $\omega\ge (2\pi)10^5\chi$ at $N=1250$,
which implies an equally large Rabi frequency $\Omega_0 \sim \omega$
of about $(2\pi)\,40$\,kHz using $\chi=(2\pi)\,0.063$\,Hz.
These are challenging conditions when $\chi$ and $N$ are both large.
Additionally, SS in this case is accompanied by a continuous nutation of the mean spin direction,
which is in contrast to the TACT model, where both the mean spin direction and
the maximal squeezing direction are fixed \cite{PhysRevA.47.5138}.

\begin{figure}
  \centering
\includegraphics[width=1.0\columnwidth]{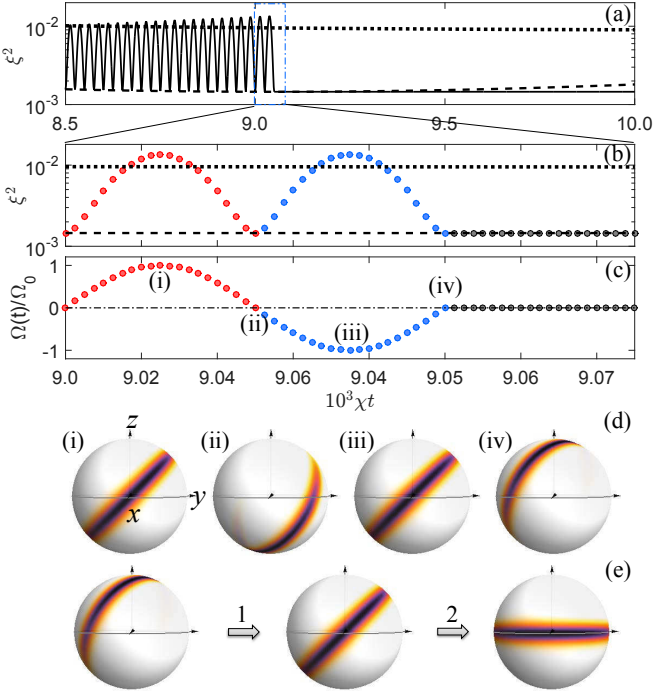}
  \caption{(Color online) Freezing of the SS governed by Hamiltonian (\ref{H_t}) with $\varphi=-\pi/2$.
  (a) Squeezing parameter $\xi^2$ as a function of time. The black solid line
  denotes the result of the modulated drive proposal aided by our protocol.
  (b)-(c) correspond to the zoomed in region marked by the blue dot-dashed rectangle in (a),
  with (b) displaying $\xi^2$ and (c) showing the scaled drive $\Omega(t)/\Omega_0$;
  (d) displays the time-dependent quasi-probability distribution $|\langle \theta ,\varphi |\psi (t)\rangle {|^2}$ at moments marked (i)-(iv)
  in (c), clearly showing the associated mean spin excursion;
  (e) illustrates our proposed manipulation to freeze the squeezing direction
  along the $z$-axis with quasi-probability distribution.
  $\omega=(2\pi)2\times 10^4\chi$, $\Omega_0=0.9057\omega$ are used, starting with the
  initial state $\exp(i\Omega_0 J_y/\omega){\left| {j,j} \right\rangle _x}$ as in Fig. \ref{fig3}.}
  \label{fig4}
\end{figure}

The difficulties associated with detecting an oscillating SS parameter and
a nutating mean spin direction can again be solved altogether if we apply our idea
to freeze SS at the optimal
[Fig. \ref{fig4}(a)]. Figure \ref{fig4}(b) and \ref{fig4}(c)
respectively show SS parameter $\xi^2$ and the modulation drive
for the time window marked by the blue dot-dashed rectangle in Fig. \ref{fig4}(a).
The coupling $\Omega(t)J_y$ from the modulated drive in Fig. \ref{fig4}(c)
continuously nutates the state around the $y$-axis,
causing its mean spin $\langle \vec J\rangle$ to oscillate
between the northern and southern hemispheres [Fig. \ref{fig4}(d)].
$\langle \vec J\rangle$ reaches its highest or lowest excursion [(i),(iii)]
when $\Omega(t) = 0$ at the moments of optimal SS [Fig. \ref{fig4}(b)].
The rotation angle of the solstice relative to the equatorial plane is estimated
to be $\int_0^{T/4} {\Omega(t) dt}  = \Omega_0 /\omega$ with $T=2\pi/\omega$ the period,
if we were to treat the $\chi J_z^2$ and $\Omega (t)J_y$ terms
as commuting with each other approximately at the large $\omega$ used.

Specifically for this case, our active rotation protocol involves
turning off the modulated drive [Fig. \ref{fig4}(c)]
when the squeezing parameter arrives at the optimal point [at $\Omega(t)=0$],
and performing a rapid state rotation to align squeezing direction to along the $z$-axis.
The rotation operation consists of two steps:
1) a rotation along the $y$-axis by an angle $\Omega_0/\omega$ to align the mean spin direction in the equatorial plane,
which results in the SS direction at $\pi/4$ relative to the $z$-axis;
and 2) a second rotation along the $-x$-axis by $\pi/4$ to rotate the SS axis to along the $z$-axis,
as illustrated in Fig. \ref{fig4}(e).
The subsequent dynamics for the state is governed by the OAT Hamiltonian $H_{\rm{OAT}}$.
The rotated squeezed state possesses a sharp distribution around $\left| {j,0} \right\rangle$,
whose squeezing parameter is frozen at the optimal point for a long time
[Fig. \ref{fig4}(a)].
If we again take the same experimentally relevant value of
$\chi\sim (2\pi)\,0.063$\,Hz, the Heisenberg limit can still be reached
even with $\omega$ and $\Omega_0$ reduced by five times to $\sim (2\pi)\,8$\,kHz.

In conclusion, we propose a scheme to freeze SS dynamics at
theoretically determined optimal moments to achieve persistent (stationary) maximum SS.
When applied to the two proposals \cite{PhysRevLett.107.013601,huang2014two} for
TACT SS at the Heisenberg limit, our scheme
 significantly relaxes stringent experimental requirements.
The number of pulses required are drastically reduced
in the repeated pulse proposal \cite{PhysRevLett.107.013601},
and modulation frequency and amplitude in the modulated drive proposal \cite{huang2014two} are also significantly reduced.
The conditions for implementing our idea seem readily achievable in
atomic SS laboratories. We believe
its experimental realization will greatly advance the pursuit of ever increasing level of SS \cite{bohnet2014reduced}.

This work is supported by the
MOST (Grant No.~2013CB922004 and No.~2014CB921403) of the National Key Basic Research Program of
China, and by NSFC (No.~91121005, No.~11374176, and No.~11328404).

\appendix
\section{The derivation of Eq. (4) in the main text}
The derivation of Eq. (4) in the main text is presented here
starting with a system described by Hamiltonian
\begin{eqnarray}\label{H}
H(t) = \chi J_z^2 + \Omega_0 \cos(\omega t+\varphi)J_y.\notag
\end{eqnarray}
A unitary transformation
$U = \exp \left[ { - i\theta \left( t \right){J_y}} \right]$
with
$\theta \left( t \right) = \int_0^t {\Omega_0 \cos \left( {\omega t_1 + \varphi } \right)d{t_1}}$,
eliminates the time-dependent coupling and reduces the Hamiltonian to
\begin{eqnarray}\label{Ht}
H &=& {U^\dag }HU - i{U^\dag }\dot U \notag\\
  &=& {R_y}\left( \varphi  \right)\chi {\left[ {{J_z}\cos {\theta _1}\left( t \right) - {J_x}\sin {\theta _1}\left( t \right)} \right]^2}{R_y}\left( { - \varphi } \right), \hskip 24pt
\end{eqnarray}
where ${R_y}\left( \varphi  \right) = \exp \left[ { - i\frac{\Omega_0 }{\omega }\sin \varphi {J_y}} \right]$ and ${\theta _1}\left( t \right) = \frac{\Omega_0 }{\omega }\sin \left( {\omega t + \varphi } \right)$. Making use of
\begin{eqnarray}\label{cos}
\cos \left[ {x\cos \left( {\omega t} \right)} \right] &=& {J_0}\left( x \right) + 2\sum_{k > 0} {{{\left( { - 1} \right)}^k}{J_{2k}}\left( x \right)\cos \left( {2k\omega t} \right)} ,\notag\\
\sin \left[ {x\cos \left( {\omega t} \right)} \right] &=& 2\sum_{k > 0}
{{{\left( { - 1} \right)}^k}{J_{2k - 1}}\left( x \right)\cos [ {\left( {2k - 1} \right)\omega t}]} ,\nonumber
\end{eqnarray}
and neglecting high-frequency oscillating terms in the high frequency approximation limit $\omega \gg \chi N$, we have
\begin{eqnarray}\label{appx}
{\cos ^2}{\theta _1}\left( t \right)
 &\simeq & [{1 + {J_0}\left( {2\Omega_0 /\omega } \right)}]/{2},\notag\\
{\sin ^2}{\theta _1}\left( t \right)
&\simeq &[{1 - {J_0}\left( {2\Omega_0 /\omega } \right)}]/{2},\notag\\
\sin {\theta _1}\left( t \right)\cos {\theta _1}\left( t \right) &\simeq& 0.
\end{eqnarray}
Upon substituting Eq. (\ref{appx}) into Eq. (\ref{Ht}), we arrive at
\begin{eqnarray}
{H_{{\rm{eff}}}} = \chi {R_y}\left( \varphi  \right)\left[ {{\alpha _0}J_z^2 + \left( {1 - {\alpha _0}} \right)J_x^2} \right]{R_y}\left( { - \varphi } \right),\notag
\end{eqnarray}
with ${\alpha_0} = \left[ {1 + {J_0}\left( {2\Omega_0 /\omega } \right)} \right]/2$.

%

\end{document}